\documentclass[aps,reprint,superscriptaddress]{revtex4-1}
\usepackage{amsmath, amssymb}
\usepackage{graphicx}
\usepackage{bm}
\usepackage{color}
\usepackage{dcolumn}
\usepackage[colorlinks=true,citecolor=blue]{hyperref}

\def\be{\begin{equation}}
\def\ee{\end{equation}}

\def\kv{{\bm k}}
\def\qv{{ \bm q}}  

\def\rv{{\bm r}}

\begin{document}
 
\title{Interplay of interlayer pairing and many-body screening in a bilayer of dipolar fermions}
\author{Azadeh Mazloom}
\affiliation{Department of Physics, Institute for Advanced Studies in Basic Science (IASBS), Zanjan 45137-66731, Iran}
\author{Saeed H. Abedinpour}
\affiliation{Department of Physics, Institute for Advanced Studies in Basic Science (IASBS), Zanjan 45137-66731, Iran}
\affiliation{Research Center for Basic Sciences \& Modern Technologies (RBST),  Institute for Advanced Studies in Basic Sciences (IASBS), Zanjan 45137-66731, Iran}

\begin{abstract}
In a bilayer system of fermionic dipoles, a full control over the strength of the attractive interactions between two layers leads to the BCS-BEC crossover.
Here, using the BCS mean field theory, we study such a crossover in symmetric bilayers of ultracold dipolar fermions with their dipole moments being perpendicular to layers. In particular, we investigate how the pairing between two layers and the many-body screening of interlayer interaction affect each other. We compare results for pairings obtained with three different approximations for the interlayer interactions namely, bare dipole-dipole interaction, the random-phase approximation for screening obtained in the normal phase, and the self-consistent superfluid phase screening within the random-phase approximation.  
We find that at weak couplings the screening further suppresses the pairing while at strong couplings, the screening would be suppressed due to the pairing gap in the quasi-particle spectrum. Therefore a self-consistent treatment of both screening and pairing on equal footings is necessary for obtaining a correct picture of the phase diagram and order parameter at both small and large layer spacings and densities.
We also notice that the highly speculated density-wave instability in bilayers with the parallel polarization of dipoles in two layers is simply an artifact of the incorrect screening scheme.    
\end{abstract}

\maketitle
 
\section{Introduction}
In a two-component fermionic system with attractive inter-component interactions, there exists a continuous state evolution from a condensate of weakly coupled particles, known as the Bardeen-Cooper-Schrieffer (BCS) state, to a condensate of tightly bounded particles, i.e., Bose-Einstein condensate (BEC), through increasing the strength of the attractive interaction~\cite{Randeria,Blattone}. 
This smooth evolution is referred to as BCS-BEC crossover, which was first investigated in excitonic systems~\cite{Moskalenko,Blatttwo,Keldysh}. In spite of enormous theoretical proposals for the observation of this crossover in platforms such as GaAs-AlGaAs heterostructures, double-layer graphene, and hybrid structures of graphene-GaAs~\cite{Sivan,Croxall,Seamons,Gorbachev,Mink,Min,Zhang,Perali,Neilson,Abergel,Zarenia14,Dell,Zarenia16}, there is yet no clear experimental evidence of interlayer superfluidity in such condensed matter structures~\cite{Neilson}. To hold out hope for the experimental realization, one has to make screening of the interlayer interactions very weak~\cite{Neilson,Abergel,Zarenia14,Dell,Zarenia16}. On the other hand, the requirement for a very clean system is felt~\cite{Abergel}.

Progress in cooling and trapping atomic gases in the recent decades, together with the immense tune-ability of the inter-particle interactions through the Feshbach resonance have provided a revolutionary route towards the better investigation and understanding of the condensed matter systems. The experimental observation of BCS-BEC crossover in ultracold atomic gases~\cite{Timmermans,Regal,Zwierlein,Kinast,Bourdel,Chin,Ketterle} might be a paramount example of this. 
Furthermore, recent successful achievements of quantum degenerate state of polar molecules and magnetic atoms~\cite{Griesmaier,Deiglmayr,Ni,Takekoshi,lu_prl2012, aikawa_prl2012, rvachov_prl2017} have offered alternative scenarios, in which the long-range and anisotropic characteristic of the dipole-dipole interaction (DDI), as well as its high controllability via external fields,  provided new avenues to the observation of topological states~\cite{Manmana,Xu}, density waves~\cite{Bruun,Yamaguchi, Sun, Sieberer, Parish, Zyl, Babadi, Wu, Zinner,Block, marchetti_prb2013,callegari_prb2017}, and exotic superfluidity~\cite{Baarsma,Abedinpour,Lee,Boudjem}. 
For instance, engineering layered structures of dipolar fermions offer a very powerful platform to study interlayer superfluidity, thanks to the attractive component of the interlayer interactions. More importantly, layered structures provide lower inelastic losses and chemical recombinations than bulk ones. This new paradigm has been extensively studied using different analytical and numerical methods~\cite{Shih,Armstrong,Miranda,Zoller,Zinnertwo,Matveeva,Pikovski}.
 
One of the key questions for the observation of BCS-BEC crossover is its stability against the many-body correlation between particles, which can dramatically affect the properties of interlayer pairing. This is in particular very important in condensed matter systems where the interaction between electrons is long range~\cite{Neilson,Perali}. 

\begin{figure}
 \includegraphics[width=0.3\textwidth]{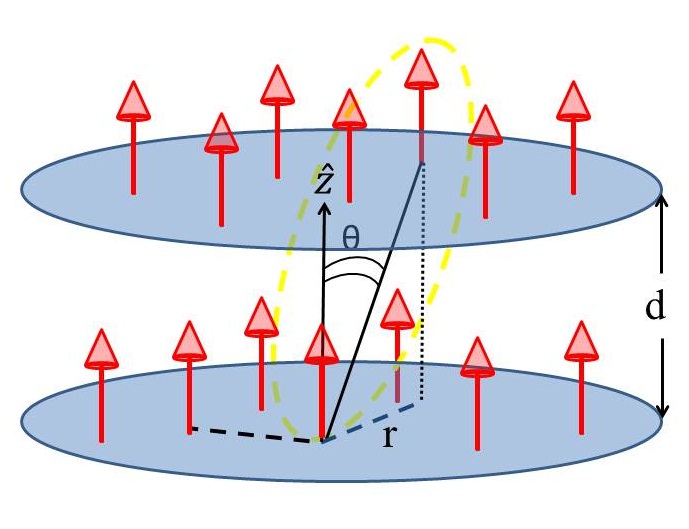}
 \caption{A bilayer system of parallel dipoles with the interlayer separation $d$. The in-plane distance between dipoles $r$ is shown. Two dipoles belonging to different layers attract each other if the angle between them $\theta$ is less than $\textrm{arctan}(\sqrt{2})\approx 54.76^\circ$.}
 \label{fig:bilayer}
\end{figure}

Here, we consider an equally populated bilayer system of identical dipolar fermions, whose dipoles are oriented perpendicular to the layers and the polarization of dipoles in two layers is in the same direction (see, Fig.~\ref{fig:bilayer}).

At the level of mean field approximation for such systems, it has been shown that~\cite{Pikovski} the superfluid-normal phase transition temperature is higher than that of the ultracold gases with contact interaction. As the mean field approach for 2D systems is only reliable at zero temperature, a beyond mean-field study has been performed in Ref.~\cite{Zoller} at non-zero temperature, which shows that many-body effects decrease (increase) the transition temperature on the BEC (BCS) side of the crossover. 
The phase diagram of dipolar bilayers has been also investigated within both BCS and strong coupling approximations at zero and finite temperatures by Zinner \textit{et al.}~\cite{Zinnertwo}.

Our prime aim here is to examine BCS-BEC crossover at \textit{zero temperature} including the effects of intra-layer screening on the interaction between dipoles of different layers. To do this, we first write down the Hamiltonian of the system and keep only the s-wave paring between particles of different layers. The BCS mean field theory enables us to obtain an equation for the superfluid gap function, which should be solved self-consistently together with the number equation to assure the fixed number of dipoles in each layer. In order to solve these equations, one needs to first specify the interlayer interaction. We employ three different approximations for it. In the first approach, we use the bare (i.e., unscreened) interlayer interaction. In the following, we will call it the unscreened (US) method. In the second approximation, the interaction between two particles belonging to different layers is screened within the random-phase approximation (RPA), while the system is considered to be in the normal state. In other words, the dielectric function of a normal bilayer is used to screen the interlayer interaction, then the superfluid gap function is obtained with this screened interaction. We will refer to this approximation as the normal-phase screening (NS).  And finally, in our most elaborate approximation, we consider that the bilayer system is in the superfluid phase and use its dielectric function to screen the inter-layer interaction and obtain the gap function. As the dielectric function itself depends on the gap parameter, screening and gap equations should be solved self-consistently. This third approximation will be referred to as the superfluid-phase screening (SS). 
 
In this paper, we compare the above mentioned three different schemes for interlayer interaction and quantify the interplay between many-body screening and the pairing between two layers. We obtain the phase diagram of bilayer system within these three approximations, where we also explore the possibility of density-wave instability (DWI), originating from the singularities of the density-density response function of the system. 
The issue of DWI in layered structures have attracted a lot of interest over the last three decades. 
It has been suggested that the enhanced correlations in layered geometries would make the observation of DWI and Wigner crystallization more feasible in electron-electron and electron-hole bilayers embedded in semiconductor heterostructure~\cite{neilson_prl91,Szymanski94,Swierkowski96,Neilson93,Moudgil02}.
In dipolar systems, the anisotropic form of the dipole-dipole interaction would in general facilitate inhomogeneous density phases, even in the single layer structures~\cite{Bruun,Yamaguchi, Sun, Sieberer, Parish}. DWI in bilayers and multilayers of dipolar fermions have been theoretically envisioned as well~\cite{Babadi, Wu, Zinner,Block, marchetti_prb2013,callegari_prb2017}, but its competition with interlayer pairing in layered structures, in particular when the dipoles are aligned perpendicular to the plane and therefore the interaction is totally isotropic has not yet well understood.
  
The outline of the rest of this paper is as follows. In Sec.~\ref{model} we present our model based on the BCS mean-field theory and find the equations of the pairing gap and the particle density of the system. Sec.~\ref{RPA} introduces our different screening approaches and discusses how to explore the possible density-wave instabilities of the system. Sec.~\ref{result} contains our numerical results illustrating the interlayer interaction and the effect of many-body screening on the BCS-BEC crossover and on the condensate fraction within different screening scenarios. Finally, our conclusions are presented in Sec.~\ref{conclusion}, where we also discuss the experimental relevance of our findings. The density-density linear response function in the superfluid state is summarized in the Appendix.

\section{Model and formalism}\label{model}
In this section, we first present the model Hamiltonian describing a bilayer of fermions interacting via the dipole-dipole interaction. We then apply the BCS mean field theory to simplify the Hamiltonian into a single particle problem, which enables one to obtain the pairing gap and the chemical potential and thereby the condensate fraction of the system.
 
 \subsection{Model Hamiltonian}
The Hamiltonian below describes a bilayer system of dipoles which are aligned perpendicular to the layers
\be\label{eq:Hamil}
\begin{split}
H&=\sum_{\kv} \xi_k^a a_\kv^\dagger a_\kv+\sum_{k} \xi_k^b b_\kv^\dagger b_\kv 
+\frac{1}{2A}\sum_\qv V_{\rm S}(q) \rho^a_\qv  \rho^a_{-\qv}  \\
&+\frac{1}{2A}\sum_\qv V_{\rm S}(q) \rho^b_\qv  \rho^b_{-\qv} 
+\frac{1}{A}\sum_\qv V_{\rm D}(q) \rho^a_\qv  \rho^b_{-\qv}  .
\end{split}
\ee
Here, the operators $a_\kv (a_\kv^\dagger)$ and $b_\kv (b_\kv^\dagger)$ are annihilation (creation) operators for dipoles in layers a and b, respectively. $\xi^{a(b)}_k=\hbar^2k^2/(2 m^{a(b)})-\mu^{a(b)}$ is the single particle energy of layer a (b) with respect to the chemical potential of the corresponding layer $\mu^{a(b)}$. 
We will consider here only symmetric bilayers, therefore $m_{a(b)}=m$ and $\mu_{a(b)}=\mu$. Third and fourth terms in  Hamiltonian~\eqref{eq:Hamil} are intralayer interactions in each layer and the last term is the interlayer interaction, with $\rho^{a}_\qv=\sum_{\kv} a^\dagger_{\kv+\qv}a_\kv$ and $\rho^{b}_\qv=\sum_{\kv} b^\dagger_{\kv+\qv}b_\kv$ being the density operators of each layer, and $A$ is the sample area.
The same layer $V_{\rm S}(r)$ and different layers $V_{\rm D}(r)$ interactions in the real space are written as~\cite{Boronat}
\begin{equation}\label{eq:potentials}
V_{\rm S}(r)=\frac{C_{\rm dd}}{4\pi}\frac{1}{r^3},
\end{equation}
and
\begin{equation}\label{eq:potentiald}
V_{\rm D}(r)=\frac{C_{\rm dd}}{4\pi}\frac{r^2-2d^2}{(r^2+d^2)^\frac{5}{2}},
\end{equation}
where $C_{\rm dd}$ is the dipole-dipole coupling constant and depends on the nature of dipole moments, $r$ is the in-plane distance between two dipoles, and the separation of two layers is indicated by $d$. 
Note that the bare intralayer interaction is purely repulsive, while the bare interlayer interaction is attractive for small in-plane separations i.e., $r \le \sqrt{2} d$ and becomes repulsive at larger separations. This introduces a critical angle of separation $\theta_c=\arctan(\sqrt{2})$ (see, Fig.~\ref{fig:bilayer}).

The corresponding Fourier transforms of intralayer and interlayer potentials can be obtained as~\cite{Block}
\begin{equation}\label{eq:spoten}
V_{\rm S}(q)=\frac{C_{\rm dd}}{4}[\frac{8}{3\sqrt{2\pi}w}-2qe^{q^2w^2/2}\mathrm{erfc}(\frac{qw}{\sqrt{2}})],
\end{equation}
and
\begin{equation}\label{eq:interpoten}
V_{\rm D}(q)=-\frac{C_{\rm dd}}{2}qe^{-qd}.
\end{equation}
Here, $\mathrm{erfc}$ is the complementary error function and $w$ is a short distance cut-off introduced to heal the divergence of the Fourier transform of $V_{\rm S}(r)$~\cite{Block}.  

\subsection{BCS mean field formalism}
As the layer indices could be regarded as the pseudo-spin degrees of freedom, one can use the standard BCS mean-field approximation to reduce the many body Hamiltonian~\eqref{eq:Hamil} to a tractable single particle problem. For the sake of simplicity, we neglect the intralayer interactions in the mean field decoupling, as their primary effect would be the Hartree-Fock renormalization of the single particle energies. However, we should note that the effects of intralayer interactions in the many-body screening of the interlayer interaction would be accounted for later on. 
We keep only the s-wave pairing between particles of different layers, which is the dominant pairing symmetry at low energies. Defining the gap function as
\be
\Delta_\kv=-\frac{1}{A}\sum_{\kv^\prime}V_{\rm D}(\kv-\kv') \langle b_{-\kv^\prime}a_{\kv^\prime} \rangle,
\ee
the mean field Hamiltonian could be written in the matrix form as 
\be\label{H_MF_matrix}
H^\textrm{MF}=
\sum_{\kv}\begin{pmatrix}
a^\dagger_\kv& b_{-\kv}
\end{pmatrix}
\begin{pmatrix}
\xi^a_k& -\Delta_k\\
-\Delta^\ast_k& -\xi^b_k
\end{pmatrix} 
\begin{pmatrix}
a_\kv\\
b^\dagger_{-\kv}
\end{pmatrix}.
\ee
With the help of the Bogoliobov transformations one finds the excitation energies of Hamiltonian~\eqref{H_MF_matrix} as $E^\pm_k=\pm E_k$, where $  E_k = \sqrt{\xi^2_k+\Delta_k^2}$.
It is evident from the excitation spectrum that, contrary to the case of imbalanced bilayers of dipoles~\cite{Abedinpour}, the system  is always gapped in the superfluid state. Minimizing the free energy with respect to the order parameter, one obtains the gap equation as 
\begin{equation} \label{eq:gap}
\Delta_k=-\frac{1}{2A} \sum_{\kv^\prime} V_{\rm D}(\kv-\kv^\prime)\frac{\Delta_{k^\prime}}{E_{k^\prime}} 
\tanh(\frac{\beta E_{k^\prime}}{2}),
\end{equation}
where $\beta=1/(k_{\rm B} T)$ is the inverse temperature. Fixed density of dipoles in each layer will give the number equation that complements the gap equation to investigate the states of system. For our symmetric system (i.e., $n^a=n^b=n$), we will have 
\begin{equation}\label{eq:chem}
n= \frac{1}{A}\sum_{\kv} a^\dagger_\kv a_\kv
= \frac{1}{2A} \sum_{\kv} \left[1-\frac{\xi_k}{E_k}\tanh(\frac{\beta E_k}{2})\right].
\end{equation}
In order to provide a description of the ground state of the system, the interlayer interaction $V_{\rm D}$ should be specified in the first place (see, next section). Afterward, a self consistent solution of the gap and number equations~\eqref{eq:gap}-\eqref{eq:chem} will give all requirements to describe the system at the mean field level. 
Once the gap function and the chemical potential are known, we are in a position to calculate the condensate fraction of the system, which for a uniform system is given by~\cite{Dukelsky, Parola}
\begin{equation}
\frac{n_0}{n}=\frac{1}{4nA} \sum_{\kv} \frac{\Delta_k^2}{E_k^2} \tanh^2(\frac{\beta E_k}{2}),
\end{equation}
where $n_0$ is the density of particles in the condensate state.

\section{Screening the interlayer potential within the random phase approximation}\label{RPA}
Although in deriving the gap equation \eqref{eq:gap}, we have ignored the intralayer interactions, their effects through the many-body screening of the interlayer potential $V_{\rm D}$ cannot be underestimated.   
In fact, the details of the approximations one uses to include these effects can qualitatively alter the final results~\cite{Neilson}. 
In this section we aim to include the effects of screening into our formalism, utilizing the previously introduced NS and SS approaches. As interlayer interactions in both approaches are screened within the RPA, this section starts with a review of it applicable to a double layer configuration. Afterward, we present the formalisms of the NS and SS schemes in subsections ~\ref{sec:normal} and \ref{sec:superfluid}. 

\subsection{Random phase approximation}\label{sec:Random}
The effective interaction matrix within the RPA is~\cite{Giuliani_and_Vignale}
\begin{equation}
 W(q,\omega) = \left[1+V(q) \chi^{RPA}(q,\omega)\right]V(q),
\end{equation}
where $V(q)$ is the matrix of the bare interactions 
\begin{equation}\label{eq:bare}
V_q
=\begin{pmatrix}
V_{\rm S}(q) & V_{\rm D}(q)  \\
V_{\rm D}(q) & V_{\rm S}(q)
\end{pmatrix},
\end{equation}
and $\chi^{\rm RPA}(q,\omega)$ is the matrix of
density-density response functions in the RPA  is written as
\begin{equation}\label{eq:rpa_matrix}
  \chi^{\rm RPA}(q,\omega)=[1-V(q)\Pi(q,\omega)]^{-1}\Pi(q,\omega).
\end{equation}
Here, $\Pi(q,\omega)$ is the matrix of non-interacting density-density response function, which for symmetric bilayers reads 
\begin{equation}
\Pi(q,\omega)
=\begin{pmatrix}
 \Pi_{\rm S}(q,\omega) & \Pi_{\rm D}(q,\omega) \\
  \Pi_{\rm D}(q,\omega) & \Pi_{\rm S}(q,\omega)
 \end{pmatrix}.
\end{equation}
We can write the eigenvalues of the RPA density-density response matrix~\eqref{eq:rpa_matrix} as
\begin{equation}\label{eq:chi_pm}
\chi^{\rm RPA}_\pm(q,\omega)=\frac{\Pi_\pm(q,\omega)}{1-V_\pm(q)\Pi_\pm(q,\omega)}, 
\end{equation}
where the symmetric (+) and antisymmetric (-) components of the bare interaction and non-interacting response function are defined respectively as
\be
V_\pm(q)=V_{\rm S}(q) \pm V_{\rm D}(q),
\ee
and
\be
\Pi_\pm(q,\omega)=\Pi_{\rm S}(q,\omega) \pm \Pi_{\rm D}(q,\omega).
\ee
Note that the dispersions of collective modes of a bilayer system could be obtained from the poles of interacting density responses $\chi^{\rm RPA}_\pm(q,\omega)$ in Eq.~\eqref{eq:chi_pm}, while its possible singularities in the static limit (i.e., $\omega=0$) signal instabilities toward density modulated phases such as the density waves.

Finally, we can easily find the screened interactions
\begin{equation}
 W_{\rm S(D)}(q,\omega)=\frac{1}{2}\left[W_+(q,\omega) \pm W_-(q,\omega)\right],
\end{equation}
where we have defined
\begin{equation}\label{eq:Wpm}
 W_\pm(q,\omega)=\frac{V_\pm(q)}{1-V_\pm(q) \Pi_\pm(q,\omega)}.
\end{equation}
Improvements upon RPA, which is essential at strong couplings, could be done through the inclusion of the local field factors (LFF)~\cite{Giuliani_and_Vignale} in the bare inter-particle interaction matrix of Eq.~\eqref{eq:bare}. 
In the following we will use the bare interlayer interaction $V_{\rm D}(q)$, as the knowledge of interlayer LFF for dipolar interactions, especially in the presence of superfluidity is lacking, but for the intralayer interaction we will use the Hubbard LFF~\cite{Giuliani_and_Vignale}, which has proven to work well at intermediate and strong couplings for dipolar systems~\cite{iran_jltp2017, emre_unpub}. This will improve our model in two ways: i) The spurious dependence of the bare intralayer interaction potential Eq.~\eqref{eq:spoten} to the short-range cutoff $w$ will be removed, and ii) the effects of exchange hole will be partially accounted for. Therefore in the following, we will replace the bare intralayer interaction $V_{\rm S}(q)$, with
\be\label{eq:vs_hub}
\begin{split}
V^{\rm H}_{\rm S}(q)&=\left[1-G^{\rm H}(q)\right]V_{\rm S}(q)\\
&=\frac{C_{\rm dd}}{2}\left[\sqrt{k_{\rm F}^2+q^2}-q\right].
\end{split}
\ee
Here, $G^{\rm H}(q)=V_{\rm S}(\sqrt{k_{\rm F}^2+q^2})/V_{\rm S}(q)$ is used, $k_{\rm F}=\sqrt{4\pi n}$ is the Fermi wave-vector of a single component two-dimensional Fermi gas, and the $w\to 0$ limit is taken at the end.

\subsection{Screening in the normal state}\label{sec:normal}
When the bilayer system is in the normal state (NS) (i.e., $\Delta_k=0$), we have $\Pi_{\rm D}(q,\omega)=0$, while $\Pi_{\rm S}(q,\omega)$ is the Stern-Lindhard function of a two-dimensional Fermi gas, which in the static limit reads~\cite{Giuliani_and_Vignale}
\begin{equation}
 \Pi^{\rm N}_{\rm S}(q)=-\nu_0[1-\Theta(q-2k_F)\sqrt{1-(2k_F/q)^2}],
\end{equation}
where $\nu_0=m/(2\pi \hbar^2)$ is the density of states. 
In this case, for the static effective interlayer interaction we find
\begin{equation}\label{eq:VNS}
 W^{\rm N}_{\rm D}(q)=\frac{V_{\rm D}(q)}{\left[1-V_+(q) \Pi^{\rm N}_{\rm S}(q)\right]\left[1-V_-(q) \Pi^{\rm N}_{\rm S}(q)\right]}.
\end{equation}
In order to study the effect of normal screening on the pairing, one has to replace $ V_{\rm D}(\kv-\kv^\prime)$ with $ W_{\rm D}(\kv-\kv^\prime)$ in Eq.~\eqref{eq:gap}. 
  
\subsection{Screening in the superfluid state} \label{sec:superfluid}
As already mentioned, in the SS approach the superfluid gap is assumed to be finite in obtaining the screening. Therefore, the interlayer component of the non-interacting density-density response $\Pi_{\rm D}$ would be nonzero. 
Starting from the mean field Hamiltonian \eqref{H_MF_matrix}, the intralayer and interlayer components of the non-interacting susceptibility are obtained as (see, the Appendix for details)
\be\label{eq:pi_S}
\begin{split}
\Pi_{\rm S}(q)=&\frac{1}{2A}\sum_{\kv} 
  \left\{  \left(1+\frac{\xi_{\kv_-} \xi_{\kv_+}}{E_{\kv_-} E_{\kv_+}}\right)
 \frac{n_{\rm F}(E_{\kv_-})-n_{\rm F}(E_{\kv_+})}{E_{\kv_-} - E_{\kv_+}}  \right.  \\
 &  \left. -\left(1- \frac{\xi_{\kv_-} \xi_{\kv_+}}{E_{\kv_-} E_{k_+}}\right)
 \frac{1-n_{\rm F}(E_{\kv_-})-n_{\rm F}(E_{\kv_+})}{E_{\kv_-} +E_{\kv_+}}\right\},
\end{split}
\ee
and 
\be\label{pi_D}
\begin{split}
\Pi_{\rm D}(q)=-\frac{1}{2A}\sum_{\kv}  &\frac{\Delta_{\kv_-} \Delta_{\kv_+}}{E_{\kv_-} E_{\kv_+}}
 \left\{  \frac{n_{\rm F}(E_{\kv_-}) -  n_{\rm F}(E_{\kv_+})}{E_{\kv_-} - E_{\kv_+}} \right.  \\
&  \left. +  \frac{1-n_{\rm F}(E_{\kv_-})-  n_{\rm F}(E_{\kv_+})}{E_{\kv_-} + E_{\kv+}} \right\},
\end{split}
\ee
respectively where $\kv_\pm=\kv \pm \qv/2$ .
It is clear that as the screening itself is a function of the gap function $\Delta_\kv$, Eqs.~(\ref{eq:pi_S}) and \eqref{pi_D} should be solved self consistently together with the gap and number equations \eqref{eq:gap} and \eqref{eq:chem}. 

It is instructive to mention here that in the US approach, the superfluid gap is generally overestimated due to the lack of screening. On the other hand, the NS scheme normally underestimates the pairing gap. Therefore the SS, which treats both screening and pairing on equal footing is expected to provide the most trustable results. 

\section{Results and discussion}\label{result}
In this section, we present our numerical results for the BCS-BEC crossover at vanishing temperature and within the three above introduced approximations for the interlayer interaction, namely unscreened, screened within the RPA for the normal state, and screened within the RPA in the superfluid state. We also address the possibility of DWI within different approximations.
We should also note that throughout this paper, we have scaled all the lengths with $r_0=m C_{\rm dd}/(4\pi \hbar^2)$ and all the energies with $\varepsilon_0=\hbar^2/(2mr_0^2)$. Moreover, we note that at zero temperature this symmetric bilayer system is identified by two dimensionless coupling constants, the dimensionless density parameter or the intralayer coupling constant defined as $\lambda=k_{\rm F} r_0$~\cite{abedinpour_aop2014}, and the dimensionless interlayer spacing $d/r_0$.

\subsection{Screened interlayer interaction and density wave instability}\label{sec:DWI}
\begin{figure}
    \includegraphics[width=0.48\textwidth]{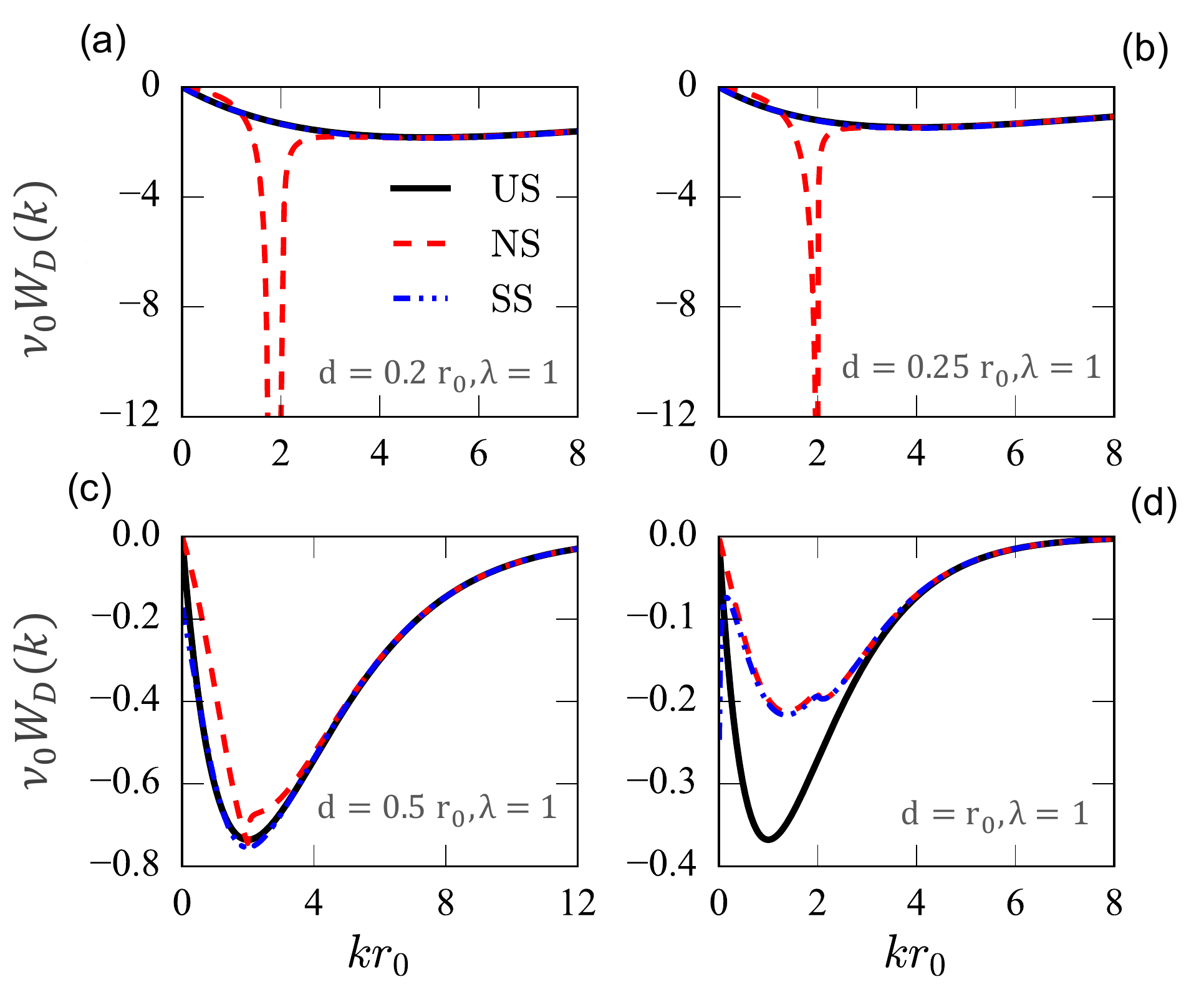}
    \caption{Comparison of the interlayer interaction obtained within different approximations for screening at $\lambda=1$ and for several values of the layer spacing. Note that for $\lambda=1$ the interlayer potential within the NS diverges below $d=0.23\, r_0$.
   \label{fig:potential}}
\end{figure}

Before turning to present our numerical results for the superfluid order parameter, we first investigate the behavior of interlayer potential $W_{\rm D}(q)$ within different approximations for screening. 
In Fig.~\ref{fig:potential} we compare the bare interlayer interaction with the screened ones (i.e., NS and SS) at a fixed layer density and for different interlayer spacings $d$. When the layer spacing is large (i.e., panel d)  as the pairing gap is expected to be negligible, both NS and SS give similar results and the screened interaction is weaker than the bare one. At intermediate and small layer separations (panels a-c), finite pairing gap suppresses the screening and therefore the SS gives similar results to the unscreened potential. When the separation between two layers becomes very small  (panel b), the normal phase screening signals a sharp peak around a specific wave vector and eventually diverges at smaller distances (panel a). 
This behavior originates from the vanishing of the dielectric function in the denominator of Eq.~\eqref{eq:Wpm} and could be a sign of a homogeneous system becoming unstable towards density modulated phases. 
Such a density wave instability is a common issue in layered structures at strong interlayer couplings. But as it is clear from the SS results, this divergence goes away once the effects of pairing are incorporated in the response functions. 
Therefore we conclude that DWI in layered structures where the attractive interlayer interaction would essentially lead to pairing gap in the spectrum is just an artifact of the inappropriate screening formalism. On the other hand, when the bare interlayer interaction is repulsive as is the case for bilayers of two-dimensional electron liquids~\cite{neilson_prl91} or dipoles with an antiparallel polarization of the dipoles in two layers~\cite{emre_unpub}, as the interlayer pairing is either absent or extremely weak, one can expect DWI at strong interlayer couplings. 

Since the self-consistent solution of the gap and particle number equations would fail due to the divergence in the normal screened interlayer potential at small layer spacings, below we analyze the criteria for this divergence and we will always stay in the homogeneous liquid region when considering the NS scheme.

Instability is identified by the poles of Eq.~(\ref{eq:chi_pm}) in the static limit. Within the NS one can get the critical distance, below which $W_+(q)$ and therefore $W_{\rm D}(q)$ have divergences. This critical spacing at weak and intermediate couplings (i.e., $\lambda \lesssim 4$) reads~\cite{emre_unpub} $2 k_{\rm F} d_c=\ln \left[\sqrt{5}/2-1+1/(2\lambda)\right]$.
 
Another quantity of interest is the average interlayer interaction
$\int\mathrm{d}\rv  V_{\rm D}(\rv)=V_{\rm D}(q=0)$,
which from Eq.~\eqref{eq:interpoten} evidently vanishes for the bare interlayer interaction  as the attractive and repulsive regions of interlayer interaction cancel each other.
The same is also true for the screened interaction in normal phase as it is clear from Eq.~\eqref{eq:VNS}.
However, at finite pairings, the average interlayer interaction could be finite. This behavior is captured only by the superfluid screening scheme.

\subsection{BSC-BEC crossover within different screening schemes}
\begin{figure}[b]
    \includegraphics[width=0.48\textwidth]{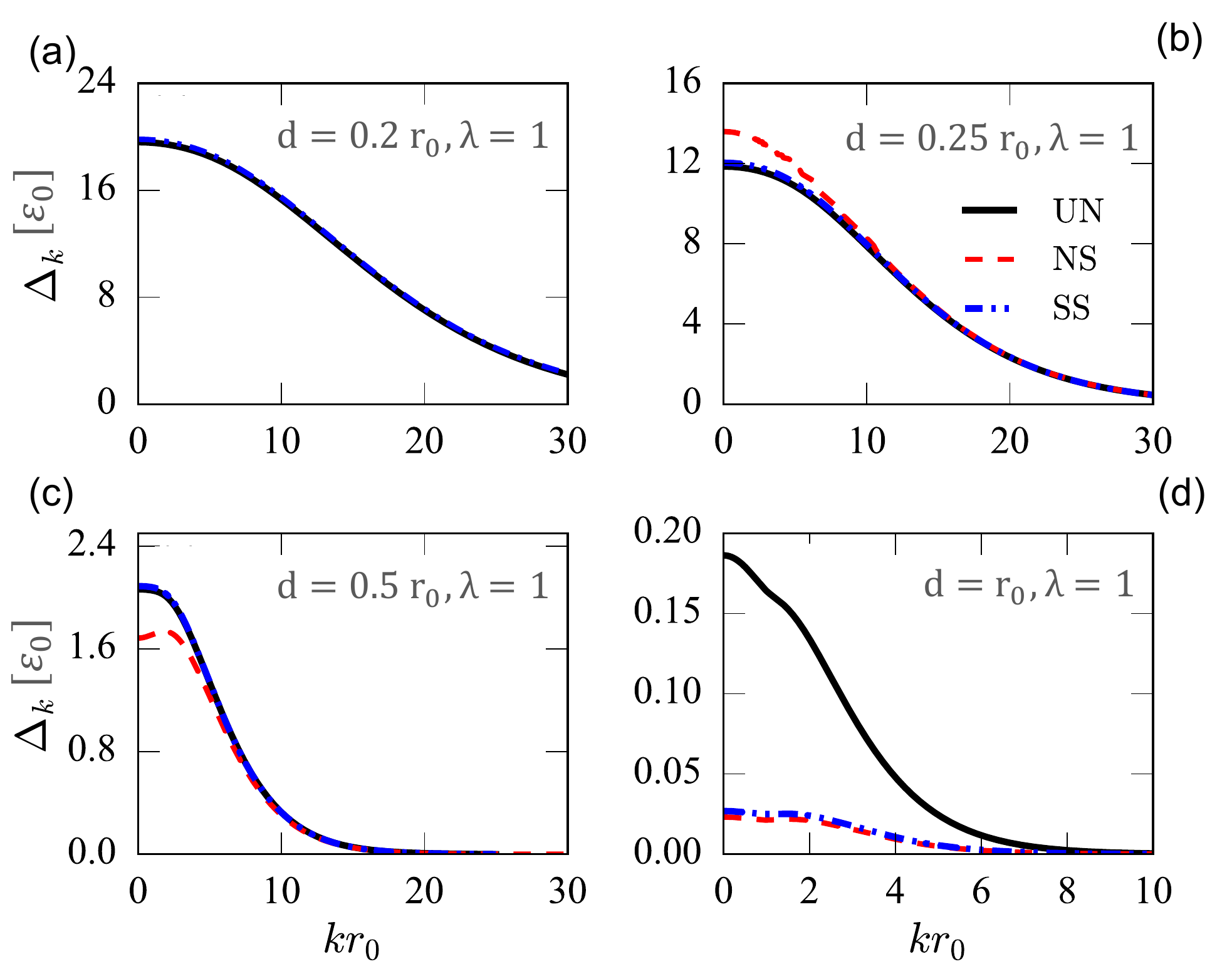}
    \caption{The wave vector dependence of the pairing gap (in units of $\varepsilon_0$) is shown for a fixed intralayer coupling constant $\lambda=1$ and for different interlayer spacings $d$. Panel (a) lies in the region where the normal screening signals density wave instability and therefore the corresponding gap function is absent there.
   }
    \label{fig:del_k}
\end{figure}

Fig.~\ref{fig:del_k} illustrates the wave vector dependence of the superfluid gap, corresponding to the same set of parameter values used in Fig.~\ref{fig:potential}. The behaviors of the pairing gaps are exactly as one expects from the behavior of interlayer potentials.
The superfluid gaps obtained from the UN and SS schemes are almost on top of each other in panels (a)-(c), as expected. In panel (a), the gap function for the NS is absent as the effective normal screened interaction diverges at small layer spacings.
The larger superfluid gap of the NS in panel (b) with respect to the bare one is just an artifact of the enhanced screened interlayer potential in the vicinity of the DWI. In panel (d) where the magnitude of the pairing gap is very small, the NS and SS schemes give similar results. We should also note that our results for the pairing gap obtained with the bare potential agree qualitatively with the ones of Zinner \textit{et al.}~\cite{Zinnertwo}.

\begin{figure}
    \includegraphics[width=0.45\textwidth]{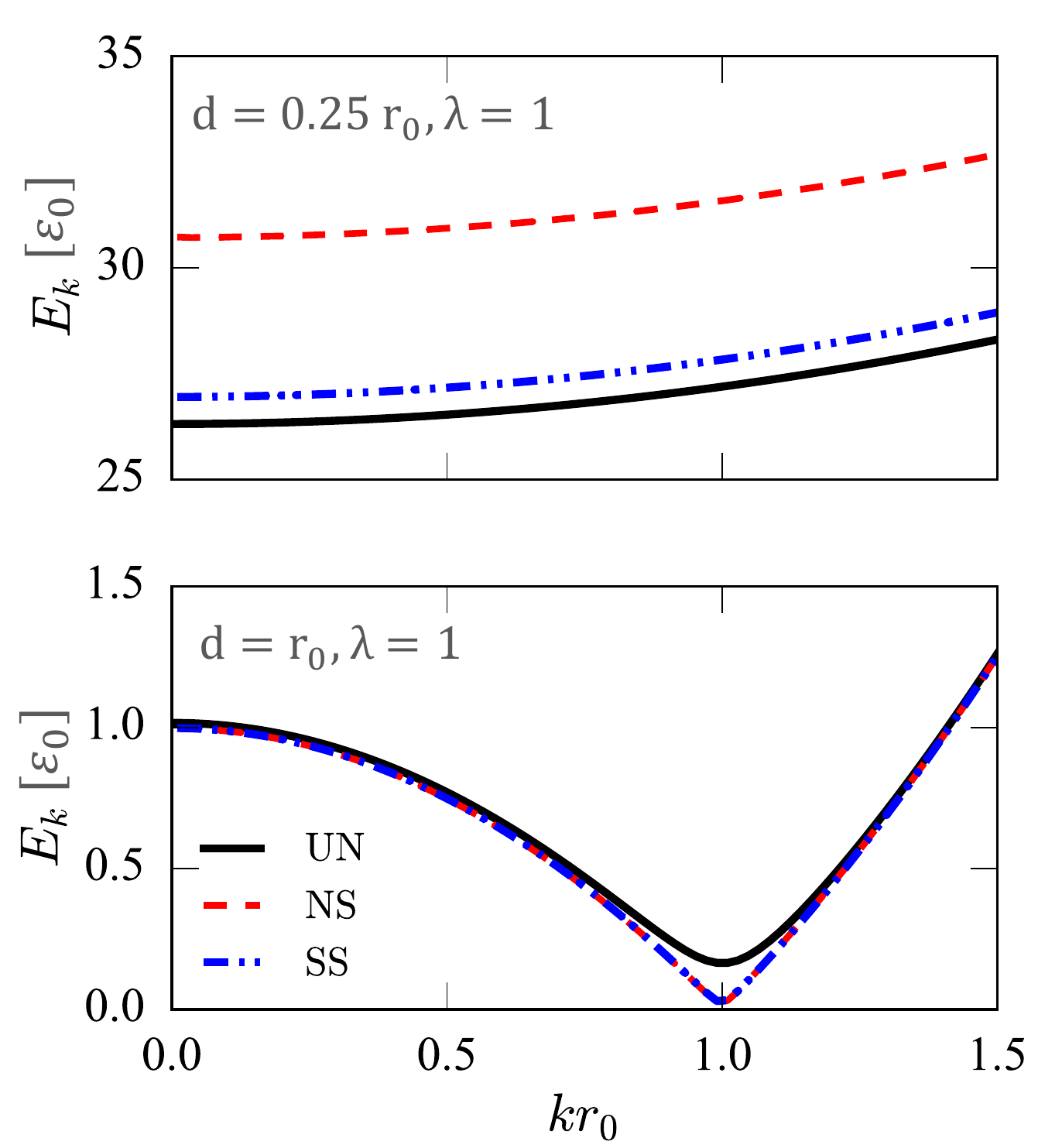}
    \caption{The quasi-particle energies (in units of $\varepsilon_0$) as functions of the wave vector obtained within different approximations for the interlayer interaction in the BEC region  (top) where $\mu<0$ and in the BCS region (bottom) where $\mu>0$. 
  }
    \label{fig:E_k}
\end{figure}

Fig.~\ref{fig:E_k} shows different behavior of the excitation energy on the BCS and BEC sides of the crossover.
In the BEC region (top panel) where the chemical potential is negative, the minimum of the excitation spectrum is at $k=0$. In the BCS region (bottom panel) the chemical potential is positive and the minimum of the excitation energy is at $kr_0\approx\sqrt{\mu/\varepsilon_0}$.

\begin{figure}
\includegraphics[width=0.48\textwidth]{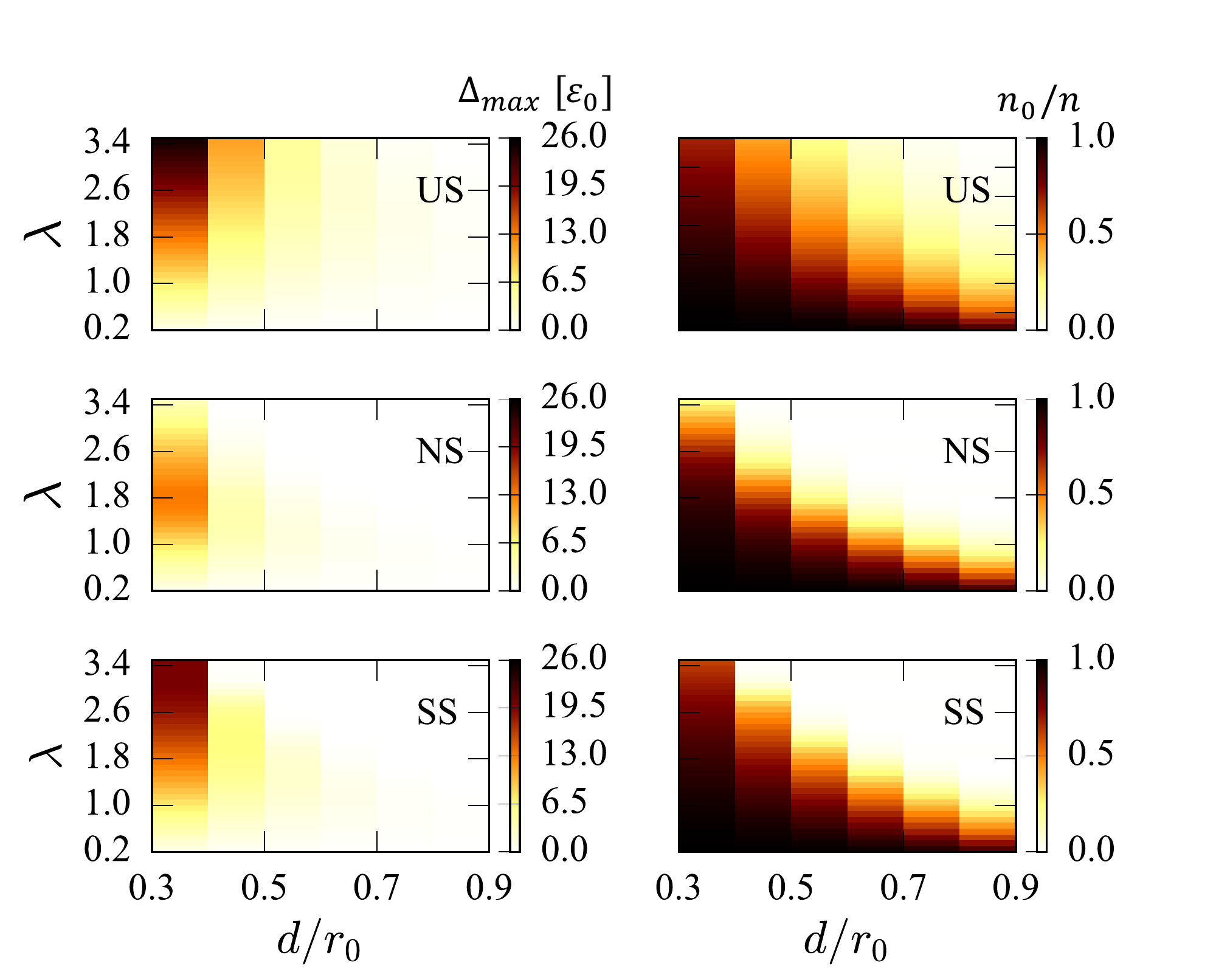}
\caption{The behaviors of the maximum values of the pairing gap (left) and the condensate fraction (right) in the $d-\lambda$ plane within three different screening schemes. 
}\label{fig:Phase}
\end{figure}

\begin{figure}
    \includegraphics[width=0.49\textwidth]{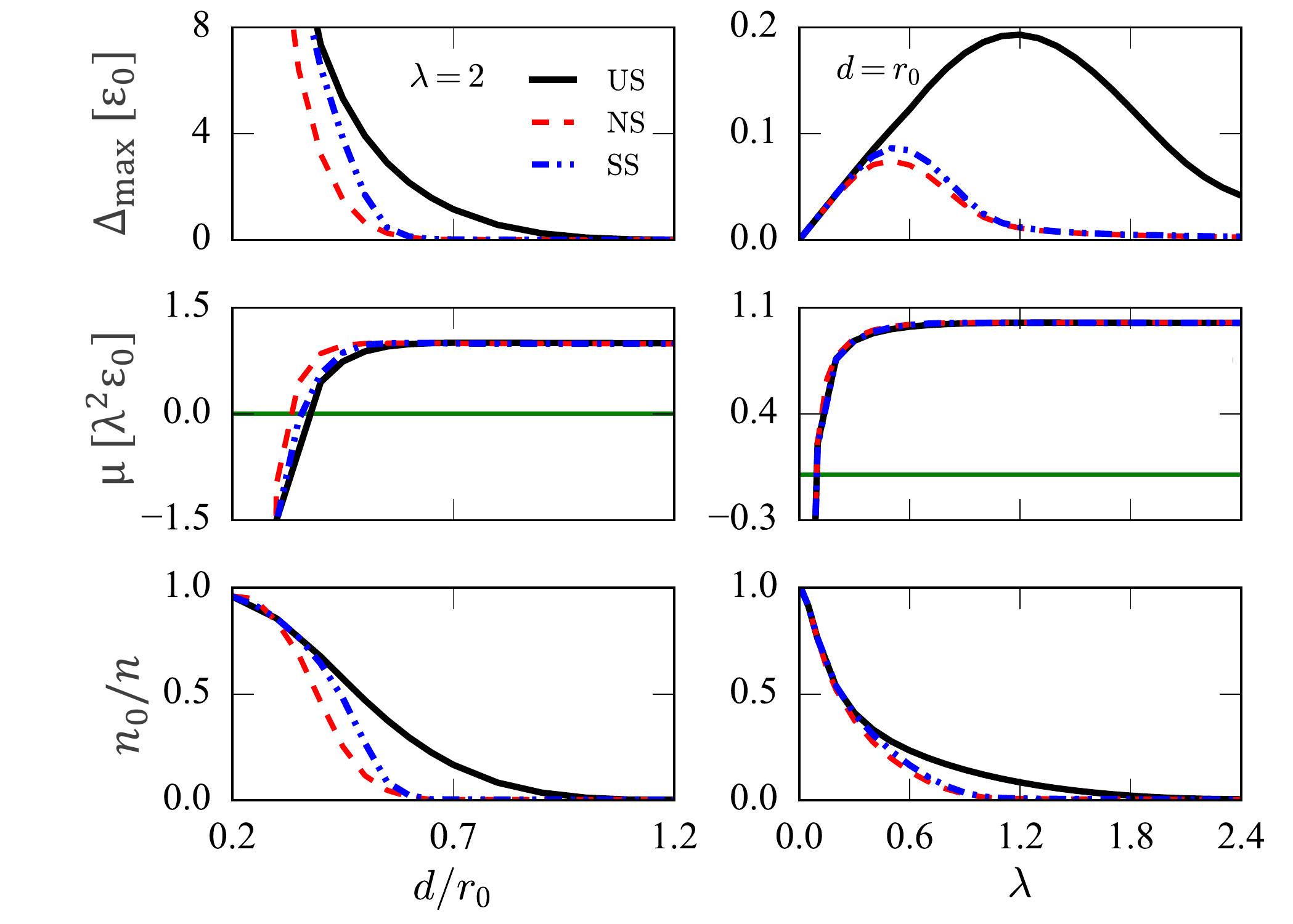}
    \caption{Illustration of the effects of many-body screening on the pairing gap (top), the chemical potential (middle), and the condensate fraction (bottom) are shown as functions of the interlayer spacing (left panels) at a fixed $\lambda$ and as functions of the intralayer coupling constant at a fixed layer spacing $d$ (right panels). Green lines in the middle panels indicate the line of zero chemical potential $\mu=0$, separating the BCS region from the BEC one.
    }
    \label{fig:del_mu_cf}
\end{figure}

We illustrate the pairing gap and the condensate fraction of a dipolar bilayer system in the ($\lambda-d$) plane, respectively in the left and right panels of Fig.~\ref{fig:Phase}. One clearly observes that at a fixed layer density, increasing the layer spacing the pairing gap decreases and eventually the system becomes normal. Incorporating the many-body screening into the formalism, the transition into normal phase moves into smaller interlayer separations. On the other hand, the superfluid gap has a non-monotonic behavior as a function of the intralayer coupling constant $\lambda$ at a fixed value of the interlayer separation. The maximum value of the pairing gap takes place at a density value which strongly depends on the screening scheme. 
Most importantly, at low-density systems superfluidity is stable to many-body screenings, whereas at high densities screening suppresses the superfluidity. 
The condensate fraction of the system decreases rapidly both with increasing the interlayer separation or the density of the system. All the described behaviors are explicitly observable in Fig.~\ref{fig:del_mu_cf}, as well. 

\begin{figure}
    \includegraphics[width=0.5\textwidth]{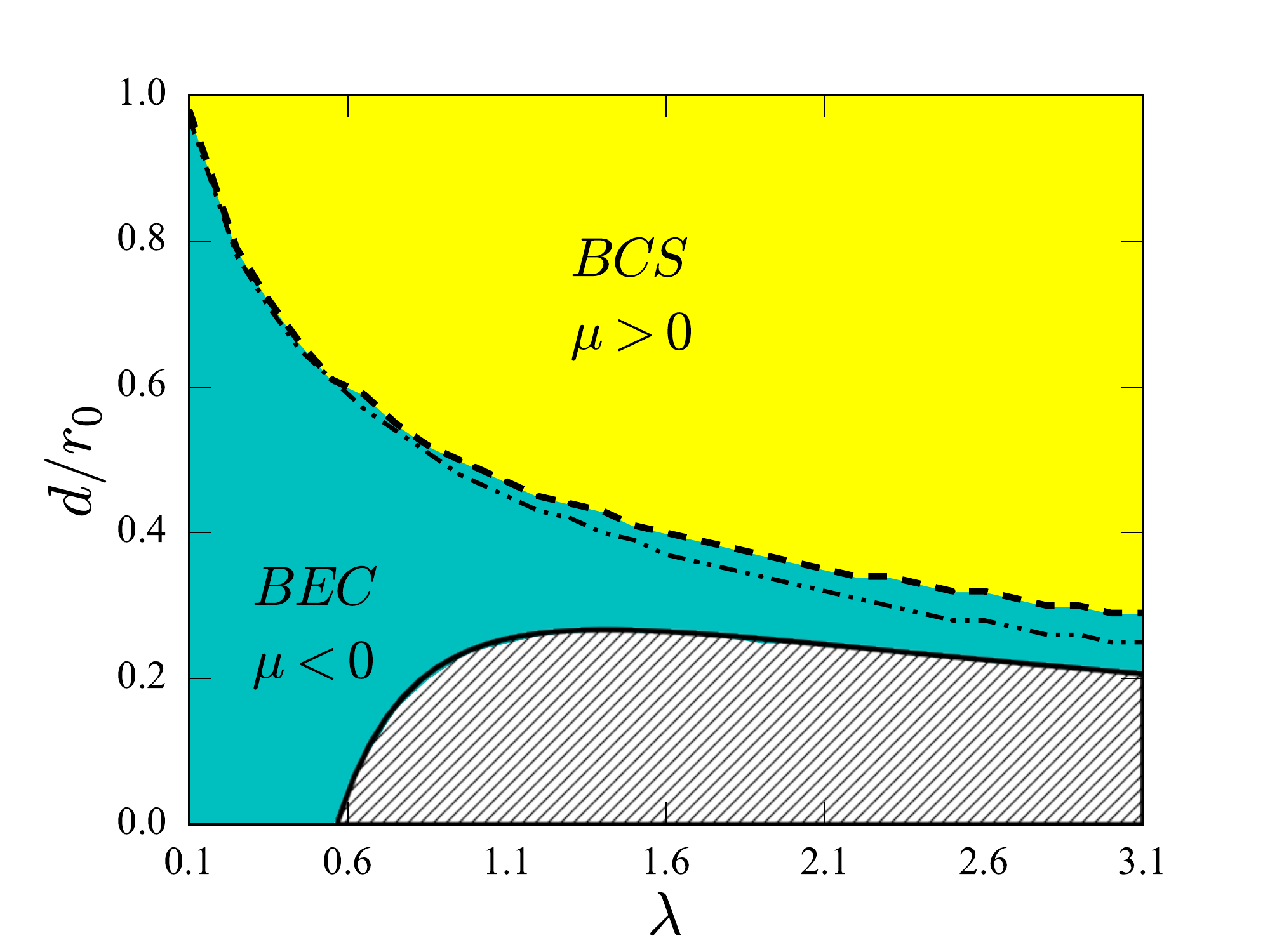}
    \caption{Phase diagram of a bilayer system of dipolar fermions as a function of the interlayer spacing $d$ and intralayer coupling constant $\lambda$. BSC  (i.e., $\mu>0$) and BEC (i.e., $\mu<0$) regions are characterized by the sign of the chemical potential. The solid and dashed lines indicate the $\mu=0$ region obtained within the US and NS schemes, respectively.
    The results of SS were indistinguishable from the unscreened one and therefore are not shown.
 The hatched area shows the region where the interlayer potential screened in the normal phase becomes divergent.}
    \label{fig:phase}
\end{figure}

In Fig.~\ref{fig:phase}, we have plotted the phase diagram of the system based on the sign of the chemical potential. Clearly, the NS slightly decreases the area of the region with negative chemical potential (i.e., the BEC side). We note that the SS screening has almost the same BCS-BEC crossover boundary as that of US interaction and therefore is not shown in the figure. The hatched area in Fig.~\ref{fig:phase} shows the region where the normal screened potential diverges, signaling the DWI.   

\section{conclusion}\label{conclusion}
In layered structures of ultracold polarized dipolar fermions, the attractive part of the interlayer interaction would be responsible for the interlayer pairing and the BCS-BEC crossover is expected through the tuning of the interlayer separation or layer densities. We have investigated the interplay between many-body screening and interlayer pairing and its impact on the BCS-BEC crossover as well as the much-speculated density wave instability. We have employed the BCS mean field theory to find the gap function for interlayer pairing, for which the effective interlayer interaction plays a prominent role. 
To include the effects of screening on the interlayer interaction, we have employed two approaches. Within the RPA, we have screened the interlayer interaction assuming that the system is either in the normal or in the superfluid phase. With a self-consistent solution of the gap and number equations, we have compared the interlayer superfluidity in the unscreened case with the NS and the SS ones. 

We have observed that within our NS scheme the bilayer system becomes unstable towards the density wave instability at small layer spacings, in agreement with many similar studies~\cite{Babadi,Parish,Zyl}. But once the effects of interlayer pairing are included in the screening this instability goes away. Therefore we conclude that the expected DWI in a symmetric bilayer system of dipolar fermions with parallel polarization is just an artifact of the poor inclusion of many-body screenings. On the other hand, if the polarization of dipoles in two layers is antiparallel, the DWI may dominate over interlayer pairing~\cite{emre_unpub}. 

Moreover, we have shown that the boundary between BCS and BEC regions in the case of unscreened interactions does not change remarkably if one screens the interlayer interactions within the SS, while the NS approach moves the boundary towards the BEC region. Our findings also indicated that at low-density systems (i.e., on the BEC side of the crossover), screening has no significant effect on the superfluid gap. In contrary at high densities (i.e., approaching the BCS regime), the screening destroys superfluidity. Furthermore, screening causes the condensate fraction to decrease more quickly, as the interlayer spacing and/or intralayer coupling constant is increased.  However, we should note that a proper description of the strong coupled BEC side requires beyond BCS mean-field techniques~\cite{Zinnertwo}.

In summary, our work has a two-fold message: First, both the BCS and BEC regimes are experimentally accessible with polar molecules such as LiCs and KRb. Second, density wave instabilities are not expected to appear in these systems.

\acknowledgements
We are grateful to D. Neilson for useful discussions.

\appendix*
\section{Derivation of the density-density response functions in the superfluid state\label{app:chi} }
Here we will illustrate how to get the density-density response functions of the BCS mean field Hamiltonian. To do this, we begin with writing the mean field Hamiltonian~\eqref{H_MF_matrix} in the Nambu notation as
\begin{equation}
 H^{\rm MF}=\sum_{\kv} \Psi_\kv^\dagger {\bar{\varepsilon}}_k \Psi_\kv,
\end{equation}
where
$\Psi^\dagger_\kv= \begin{pmatrix}  a^\dagger_\kv & b_{-\kv} \end{pmatrix}$,
and
\begin{equation}
{\bar{\varepsilon}}_\kv=
 \begin{pmatrix}
 \xi_\kv & -\Delta_\kv \\
 -\Delta^*_\kv & -\xi_\kv
 \end{pmatrix}.
\end{equation}
The total density operator with this notation is
\begin{equation}
  \rho_\qv=\sum_{\kv} \Psi_{\kv-\qv/2}^\dagger \tau^z \Psi_{\kv+\qv/2},
\end{equation}
where $\tau^z$ is the z-component of the Pauli matrix which acts on the layer degree of freedom. Layer resolved density operator for layer $\alpha$ reads
\begin{equation}
  \rho_{\qv,\alpha}=\sum_{\kv} \Psi_{\kv-\qv/2}^\dagger \tau^\alpha \Psi_{\kv+\qv/2},
\end{equation} 
where $\tau^{a(b)}=(\tau^z\pm 1)/2$.
The Matsubara Green's function matrix in the Nambu formalism is introduced as~\cite{bruus_flensberg}
\begin{equation}
{\bar{G}}(k,i\varepsilon_n)=
  \begin{pmatrix}
   G(k,i\varepsilon_n) & F(k,i\varepsilon_n) \\ F^*(k,i\varepsilon_n) & -G^*(k,i\varepsilon_n)
  \end{pmatrix}.
\end{equation}
Here, the normal and anomalous Green's functions are defined as
\begin{equation}
  G(k,i\varepsilon_n)=\sum_{\lambda=\pm 1} \frac{w_k^\lambda}{i\varepsilon_n-\lambda E_k},
\end{equation} 
and
\begin{equation}
  F(k,i\varepsilon_n)=\frac{-\Delta_k}{2E_k}\sum_{\lambda=\pm 1} \frac{\lambda}{i\varepsilon_n-\lambda E_k},
\end{equation}
respectively, where $w^\pm_k=(1\pm \xi_k/E_k)/2$ and $\varepsilon_n=(2 n+1)\pi/\beta$ with integer $n$ is a fermionic Matsubara frequency. 
The density-density response function within the single bubble diagram level reads
\begin{widetext} 
\be
\Pi_{\alpha \alpha'}(q,i\omega_n)=\frac{1}{A\beta}\sum_{\kv,i\varepsilon_n}
\mathrm{Tr}\left[{\bar{G}}(\kv_-,i\varepsilon_n)\tau^\alpha {\bar{G}}(\kv_+,i\varepsilon_n+i\omega_n)\tau^{\alpha'} \right],
\ee
where $\kv_\pm= \kv \pm \qv/2$, and $\omega_n=2 n \pi/\beta$ is a bosonic Matsubara frequency.  After some straightforward algebra we find
\be
\begin{split}
\Pi_{\rm S}(q,i\omega_n)
&=\frac{1}{A\beta}\sum_{\kv,i\varepsilon_n} G(\kv_-,i\varepsilon_n) G(\kv_+,i\varepsilon_n+i\omega_n) \\
&=\frac{1}{A}\sum_{\kv ,\lambda,\lambda^\prime} w^\lambda_{\kv_-}w^{\lambda^\prime}_{\kv_+}
\frac{n_{\rm F}(\lambda E_{\kv_-})-n_{\rm F}(\lambda^\prime E_{\kv_+})}{i\omega_n+\lambda E_{\kv_-}-\lambda^\prime E_{\kv_+}},
\end{split}
\ee
and 
\be
\begin{split}
\Pi_{\rm D}(q,i\omega_n)
&=\frac{1}{A\beta}\sum_{\kv,i\varepsilon_n} F^*(\kv_-,i\varepsilon_n) F(\kv_+,i\varepsilon_n+i\omega_n)  \\
&=-\frac{1}{A}\sum_{\kv,\lambda,\lambda^\prime} 
\left(\frac{\lambda \lambda^\prime \Delta_{\kv_-}\Delta_{\kv_+}}{4E_{\kv_-}E_{\kv_+}}\right)
 \frac{n_{\rm F}(\lambda E_{\kv_-})-n_{\rm F}(\lambda^\prime E_{\kv_+})}{i\omega_n+\lambda E_{\kv_-}-\lambda^\prime E_{\kv_+}}.
\end{split}
\ee
\end{widetext} 
In the static limit, after performing the summation over $\lambda$ and $\lambda'$, we arrive at Eqs.~(\ref{eq:pi_S}) and \eqref{pi_D}.


\end {document}